\documentclass[pre,twocolumn,showpacs,preprintnumbers]{revtex4}
\usepackage{amsmath,amssymb,latexsym}

\usepackage{graphicx}

\begin{document}

\title
{Electron-ion and ion-ion potentials
for modeling warm-dense-matter:
applications to laser-heated or shock-compressed Al and Si. 
}

\author
{
 M.W.C. Dharma-wardana}
\affiliation{
National Research Council of Canada, Ottawa, Canada, K1A 0R6
}
\email[Email address:\ ]{chandre.dharma-wardana@nrc-cnrc.gc.ca}

%
\date{\today}

\begin{abstract} 
The   pair-interactions $U_{ij}(r)$ determine the thermodynamics and
linear transport properties of matter via the 
pair-distribution functions (PDFs), i.e., $g_{ij}(r)$. Great simplicity
is achieved if $U_{ij}(r)$ could be directly used to predict material
properties via classical simulations, avoiding many-body wavefunctions.
Warm dense matter (WDM) is encountered in quasi-equilibria where the
electron temperature $T_e$ differs  from the ion temperature $T_i$, as
in laser-heated or in shock-compressed matter.  The electron PDFs
$g_{ee}(r)$ as perturbed by the ions are used to evaluate fully
non-local exchange-correlation corrections to the free energy,  using
Hydrogen as an example. Electron-ion potentials for ions with a bound
core are discussed with Al and Si as examples, for WDM with $T_e\ne
T_i$, and  valid for times shorter than the electron-ion relaxation
time. In some cases the potentials develop attractive regions, and then
become repulsive and `Yukawa-like' for higher $T_e$. These results
clarify the origin of initial phonon-hardening and rapid release. 
Pair-potentials  for shock-heated WDM  show that phonon hardening would
not occur in most such systems.  Defining meaningful quasi-equilibrium
{\it static} transport  coefficients consistent with the dynamic values 
is addressed. There seems to be no meaningful `static conductivity'
obtainable by extrapolating experimental or theoretical $\sigma(\omega,
T_i, T_e)$ to $\omega \to 0$, unless $T_i \to T_e$ as well. Illustrative
calculations of quasi-static resistivities $R(T_i,T_e)$ of laser-heated
as well as shock-heated  Aluminum and Silicon are presented
using our pseudopotentials, pair-potentials and classical integral
equations.  The quasi-static resistivities display clear differences
 in their temperature evolutions, but are not the strict
  $\omega \to 0$  limits of the dynamic values.
\end{abstract}

%
\maketitle
\section{Introduction}
Novel techniques of high-energy deposition on matter using 
high-energy short-pulse lasers as well as shock waves enable one to
produce matter in a variety of novel states. They demand theoretical
techniques adapted to many physical regimes.  Such warm-correlated
matter (WCM), and warm dense mater (WDM),
  include hot-nucleating nanocrystals we well as hot-dense radiative (HDR) 
plasmas~\cite{cimarron}. They have  applications ranging
from laser micro-machining, laser ablation~\cite{Lorazo-Lewis03},
Coulomb explosions~\cite{Lewis},  inertial-confinement fusion (ICF),
astrophysics~\cite{Militzer12} and aero-space re-entry
applications~\cite{SGA89}.  The material may be
electrons and ions in a complex mixture of 
ionization states of low-$Z$ and high-$Z$ ions, e.g., H to U.
 The temperatures could be very high and yet high
compressions could make the electrons degenerate or
 partially degenerate.
Predictions of the thermodynamics, transport, radiative, and
thermonuclear processes \cite{cimarron} pose major challenges.
Furthermore, ion temperatures $T_i$ may differ
 from the electron
temperature $T_e$, and the quasi-equilibrium
properties, relaxation and transport have to be updated within
the time-scales and energy-scales of such two-temperature
equilibria~\cite{Ng2011}.

 Modern electronic-structure
methods provide {\it in situ} density-functional
potentials 
incorporated into molecular dynamics (MD)
simulations~\cite{Car-Par85}. 
The all-electron, all-ion quantum
calculations for simple systems provide
`bench-marks', usually at low temperatures and normal
densities. Such methods become untenable at  higher-$T$
where large numbers of partially
 occupied eigenstates have to be included.
 HDR plasmas, relevant to ICF provide a
perspective of the problem. The complexity of WDM, compressions and
 temperatures involved call
for simplifications without sacrificing accuracy. Thus molecular
dynamics (MD), i.e., classical simulations, using various  effective
potentials have been a focus of some recent studies~\cite{JonesMur07,
YeZhaoZhen10}. Such methods were used in earlier work~\cite{Ross80},
often with phenomenological or `chemical' models for
pair-interactions, and in recent studies as well~\cite{HouY09}.  
Other authors have studied statistical potentials inclusive of
quantum diffraction effects~\cite{DuftyDutta} arising from integrating
out the electrons. A re-examination of potentials useful in the
non-equilibrium context is needed.

We study two types of non-equilibrium (two-temperature)
systems, viz., (i) generated by short-pulse lasers (the ion subsystem
remains virtually unchanged while $T_e$ increases during the pulse period) and
 (ii) mechanical shocks. Here the electrons remain at $T_e$ while $T_i$ and
the compressions change.  Our approach has been to construct
the charge distribution $n(r)$ around a nucleus immersed
 in the medium via the
neutral-pseudoatom (NPA)  density-functional theory (DFT) models, e.g., 
Ref.~\cite{eos95}. The $n(r)$ is used to construct a
local {\it weak} pseudopotential dependent on the density and
temperature of the ambient environment. The pseudopotentials $U_{ei}(r)$ 
provide pair-potentials $U_{ij}$ and pair-distribution functions (PDFs)
$g_{ij}(r)$.
These provide the thermodynamics and transport
properties of the system in a self-consistent manner. The dependence
of the $U_{ei}$ on the system parameters arises because the
effective core-charge $\bar{Z}$, the ionic-core radii $r_c$, etc., 
of the ion-configuration depend on the medium. 
A type of transferable pseudopotentials
is available with popular simulation software. However, they assume
frozen-core radii $r_c$ and an ionization $\bar{Z}$ consistent
with normal densities and $T=0$. They are not transferable to
highly compressed WDM states, or low electron degeneracies,
 as is well known and noted even
recently by Mazevet el al.~\cite{Mazevet07}.
A more detailed approach that does not require defining a `core'
would be to use all-electron regularized
pseudo-potentials based on  norm-conserving (NC) or projector-augmented-wave (PAW) approaches. Such methods would be strongly computer-intensive and
not useful for most WDM problems.
They could serve to provide a new set of benchmarks that are
beyond Quantum-Monte Carlo calculations, rather than
serve actual WDM calculations.

Using such NC, PAW or  `standard' 
pseudopotentials requires solving the
 many-center Schr\"{o}dinger equation, as
implemented in major codes~\cite{codes}.
The numerical simplicity needed for studying complicated WCM
systems with many ionization states and components in different
temperature states is lost. Hence we focus on
simple linear-response potentials as in, e.g., 
Ref.~\cite{eos95}. The ion subsystem can be treated using MD or
via integral equations like the
hyper-netted-chain (HNC), or the modified-HNC (MHNC~\cite{Rosen})
 method to exploit spherical symmetry. 
This works even for
quantum electrons at $T=0$ via a classical map~\cite{prl1}. 
Hence the present study is an extension 
 to non-equilibrium (two-temperature) systems generated
by short-pulse lasers or mechanical shock. The physics expressed in
terms of pair-potentials and PDFs can be directly generalized to 
deal with  $T_e \ne T_i$, i.e.,  two-temperature quasi-equilibria,
using  $g_{su}(r,T_s, T_u)$, where $s,u$ are electron or ion
subsystem labels. Such generalizations can be formally justified in
terms of the Bogolubov-Zubarev type of non-equilibrium
theory~\cite{elr0}.  Within the time scales $\tau_{ei}$ where subsystem
Hamiltonians $H_e$ and $H_i$ remain invariant, we can also justify
the use of DFT for each subsystem.  

 Since laser-pulsed heating or shock-compression
experiments begin at some reference density near room
temperature, the pseudopotentials can be checked 
against experimental liquid-metal
 properties. We
use such liquid-metal-adapted pseudopotentials, with pair
potentials constructed from finite-$T$ response functions
incorporating finite-$T$ local fields consistent with
the sum rules and finite-$T$ exchange-correlation
effects~\cite{prbexc}. 
Equilibrium and quasi-equilibrium WDM Aluminum and Si are studied
in detail. Experimentally useful quantities accessible {\it via} these
 calculations are  (a) pair-distribution functions and
 structure factors (b) thermodynamics, e.g., Hugoniots,
 subsystem free energies, etc.
(c) static and dynamic liner transport coefficients,
 (d) energy-relaxation rates
 (e) X-ray Thompson scattering and other dynamical results.
 In this paper we address
various aspects of (a)-(d) and leave (e) for a future study.

We also  ask if the electron-ion pseudopotentials $U_{ei}(r)$ or
the pair-potentials  $U_{ii}(r)$ could be approximated by a
Yukawa form $W_{ss'}$, and compare them with more detailed potentials.

Here we note that DFT or Car-Parinello methods cannot
calculate the electron-electron PDFs $g_{ee'}(r)$, where $e, e'$
are electrons (with spin indices included as needed). However, these
electron-electron PDFs may be calculated using a well-tested
classical-map technique (CHNC)~\cite{prl1, ijqc11}, inspired by
finite-temperature DFT itself. This is used in sec.~\ref{ex-corr.sec}
to obtain the fully nonlocal exchange-correlation free energy of 
electrons interacting directly with ions, rather than jellium. In 
sec.~\ref{ex-corr.sec} we use the electron-electron pair-distribution
functions $g_{ee}(\lambda, r,T)$, where $\lambda$ is a 
coupling constant,  to evaluate the difference
 between  the exchange-correlation free energy of electrons in jellium
and in hot hydrogen via a fully non-local evaluation using
a coupling constant integration of  $g_{ee}(r,T)$ in hydrogen.

 Section \ref{pseudp-sec} presents the details of the pseudopotential
and PDF calculations for ions with a core specified by the charge
$\bar{Z}(T_i)$ and the radius $r_c$. Results for
equilibrium and quasi-equilibrium Al and Si WDMs are given. 
 The section \ref{two-T-transport} on transport properties
addresses the meaning of a `static' conductivity as the frequency
independent limit of the two-temperature dynamic conductivity 
$\sigma(\omega, T_i, T_e)$.
It is argued that there may be no physical meaning in the `static conductivity'
obtained by extrapolating $\sigma(\omega, T_i, T_e)$
to $\omega \to 0$, unless $T_i \to T_e$ as well.
Calculations are presented for a possible candidate to a
  `quasi-static' resistivity of
laser-heated and shock-heated Aluminum, showing how they differ
from each other and from the $T$-dependent
equilibrium resistivity. Such results can be used to test
if the physical models presented here provide a
reasonably accurate picture of laser-pulse heated or  shock-heated
materials in the WDM regime.

\section{Non-local exchange-correlation calculations.}
\label{ex-corr.sec}
An important application of finite-$T$ DFT is to evaluate  the
exchange-correlation contribution $f_{xc}(T)$ per electron 
to the Helmholtz 
free-energy $f$ of a given electron distribution $n(r)$ at the
temperature $T$. This is usually done in the 
local-density approximation (LDA) using the 
$f_{xc}(T)$ per unit volume of the uniform electron gas (UGE)
as the input functional. 
Some authors have even approximated this with the $T=0$ 
UGE parametrization, even though good approximations to $f_{xc}$
from evaluations of finite-$T$ bubble 
diagrams~\cite{Perrot-Fxc,Batonrouge}, the
Iyetomi-Ichimaru (IYI) parametrization~\cite{Ichimaru}
as well as the $f_{xc}(T)$ from CHNC
have been available for sometime~\cite{prbexc}.
The IYI scheme and the CHNC $f_{xc}$ agree well in
typical WDM regimes of density and $T$. 

 In this section we present a fully non-local calculation of
 $f_{xc}$ at finite $T$
for a system of electrons interacting with a subsystem of protons,
using classical potentials and pair-distribution functions as the
ingredients of our calculation. The
protons are in thermal equilibrium with the electrons, and
have kinetic energy. However, no Born-Oppenheimer approximation is
invoked. The ions are a system of
classical particles and no classical map is needed.
 The proton free energy has
 a correlation contribution $ F^{(p)}_c$ which gets automatically
 evaluated through the HNC procedure. It is a highly non-local
 quantity that cannot be evaluated in the LDA
 even as a first approximation, but we do not dwell on it here.
In fact, the total free-energy given by CHNC was used to calculate a
Deuterium Hugoniot in ~\cite{Hug-H} but the nature
 of the purely electronic part $f_{xc}(r_s,T)=F_{xc}/N$,
 where N is the total number of electrons,  was not examined there
in detail. Here we high-light that aspect and
 concentrate on the electronic part of the exchange-correlation
 free energy $F_{xc}$. 

The Hamiltonian of the system and its free energy calculation
are similar to that of the electron gas,
except that the unresponsive `jellium' background is replaced by the
proton subsystem. Thus
\begin{eqnarray}
\label{h-ham.eqn}
H&=&H^0_e+H^0_p+H_{ep}+H_{ee}+H_{pp} \\
F&=&F^{mf}_e+F^{mf}_p+F_{ep}+F_{xc}+F^{(p)}_c
\end{eqnarray}
Here the superfix {\it mf} stands for `mean-field',
and all the terms have an obvious meaning. 
 Since the electron and proton densities are
identical, the electron Wigner-Seitz radius $r_s$ is also the
proton $r_s$. The classical map is needed
only for the electrons. We have used the HNC version of
the classical map~\cite{prl1}, i.e., CHNC, for this calculation.
In CHNC the quantum electrons (even at $T$=0) are treated as
a Coulomb gas at a finite temperature dependent on $r_s$ and $T$. The
electrons interact via the diffraction-corrected Coulomb potential plus a
Pauli exclusion potential. The latter accurately reproduces
 the exchange hole in 
the parallel-spin $g_{ee}(r)$~\cite{Lado}.   
The density ($\sim$ 0.8 g/cc) in this illustrative calculation is
 chosen so that we have a
fully ionized e-p gas at a high compression and $r_s=1.5$ au.
The treatment of ions with bound states
 may follow the methods for Si and Al given below.

\begin{figure}
\includegraphics*[width=8.5 cm, height=6.0 cm]{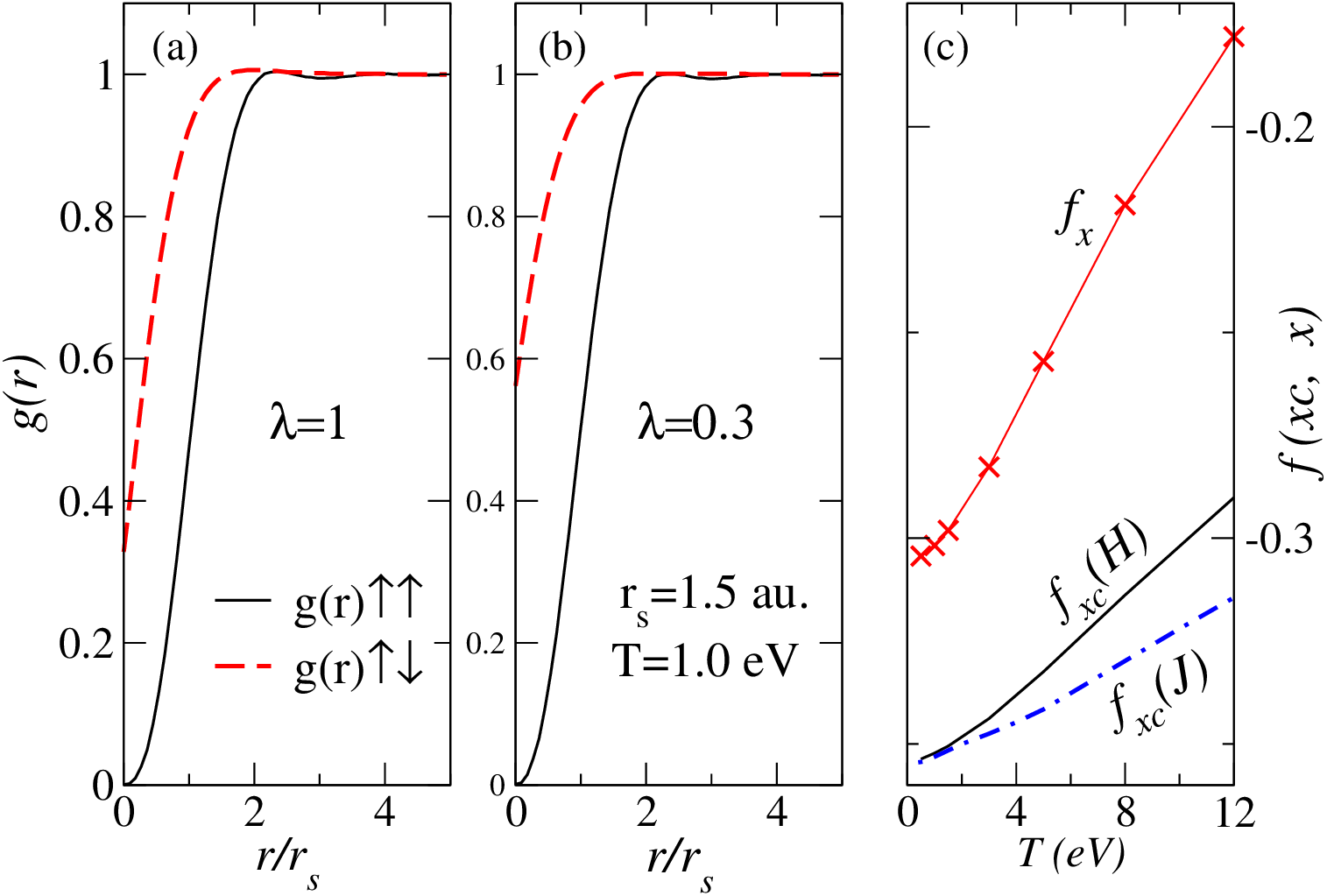}
 \caption
{(Online color) (a)The parallel and anti-parallel electron
pair-distribution functions $g_{ee}(r)$ at the full value of the Coulomb
interaction, i.e., $\lambda$ = 1 in a system containing interacting
protons and electrons  at $r_s$ = 1.5 au. and $T$ = 1 eV. (b)
The same PDFs evaluated at $\lambda$ =0.3, and for the same $r_s,T$ 
as before. (c) The non-local exchange-correlation free energy $f_{xc}$
(solid line) in the hydrogen fluid, evaluated from the coupling-constant
integration over the PDFs, compared with a similar
 calculation~\cite{prbexc} for
jellium (dash-dot line) at $r_s$ = 1.5 and $T$ = eV. 
The exchange energy $f_x$ is evaluated from the PDFs 
at $\lambda=0$.
Contributions to correlations from $g_{ep}(r)$ are not included here.} 
\label{hyd-fxc.fig}
\end{figure}

In the panels (a), (b)  of fig~\ref{hyd-fxc.fig}, we display the electron
pair-distribution functions $g_{ee}(\lambda,r, T)$ for 
values of the coupling constant $\lambda$ = 1, and 0.3.
 The Coulomb interaction $1/r$ is replaced by $\lambda/r$ with
 $\lambda$ varied from 0 to 1. The fully non-local
 exchange-correlation free energy per electron, $f_{xc}=F_{xc}/N$,
is given by an integration of $h_{ij}(r)=g_{ij}(r)-1$ over $\lambda$.
\begin{equation}
\label{adiabatic-eq}
f_{xc}=\frac{n}{2}\int_0^1\frac{d \lambda}{2\lambda}
\int\frac{\lambda d{\vec{r}}}{r}[h_{11}(r,\lambda)+h_{12}(r,\lambda)].
\end{equation}

It is noteworthy that the parallel-spin PDF, i.e.,  $g_{11}(r)$
 is hardly affected by the
value of the coupling constant.  It is dominated entirely by the Pauli
exclusion effect. Hence the correlations are mostly mediated by the
anti-parallel $g(r)$ which changes drastically with $\lambda$. The
success of the use
of an `exact exchange'   in `optimized-effective potentials' and
related methods~\cite{Krieger92} is related to this property
of $g_{11}(r)$. Spurious `self-interaction' errors are absent
in CHNC implementations or any methods that directly use a $g(r)$.
In panel (c) we present our values of
$f_{xc}(H)$, i.e., the  xc-free energy per electron in
the presence of protons,  and the corresponding quantity,
 labeled $f_{xc}(J)$ in the
absence of protons, appropriate for the finite-$T$ jellium model. The
exchange free energy $f_x$ is simply the exchange energy at $\lambda=0$,
and is the same for the system with protons (Hydrogen) or without
protons (i.e., jellium). Thus the correlation contribution is the many-body
correction beyond the jellium $f_x$ at finite-$T$. The reference density
matrix is diagonal in a set of plane-waves. The Hartree energy is zero.
 This definition of
correlation energy is different to a treatment where the electron-proton
fluid is treated as a system of hot atoms together with
correlation corrections. There the correlation energy is usually defined with
reference to a density matrix diagonal in an atomic Hartree-Fock basis. The
corrections to the energy beyond the Hartree-Fock value are deemed
the correlation energy. 
Such a basis already has electron-ion correlations in the wavefunctions
which are not plane waves. 
They are not
included in the $f_{xc}$ used here which is purely the electronic part.
  In effect, the
correlation corrections in $g_{ei}(r)$ are differently partitioned in
these different definitions, and hence care must be taken in making
comparisons between atomic-like  Hartree-Fock approaches to
  WDM~\cite{sjostrom12}, and associated discussions of finite-$T$
 correlation energies.

\section{The pseudopotentials.}
\label{pseudp-sec}
Accurate pseudopotentials are available with
standard computational packages~\cite{codes}, designed for
condensed-matter applications where finite-$T$ effects,
multiple ionizations, non-equilibrium effects etc., are uncommon. WCM
applications require accurate but simpler potentials where
 classical simulations,
linear response methods, etc., could be used instead of basis-set
diagonalizations of a Hamiltonian.  Here we consider (i) the simplest
Yukawa type potentials, and (ii) more accurate pseudopotentials from
DFT for a single nucleus immersed
in a WDM medium. Thus a quantum calculation is necessary only for a
single nucleus in a spherically symmetric environment modeled as a
`neutral pseudo atom' (NPA)~\cite{eos95}. The ion
density  is at the ion
temperature $T_i$. A DFT 
calculation at $T_e$ gives the electron density $n(r)$
around the nucleus, the effective ionic charge $\bar{Z}$, 
Kohn-Sham wavefunctions and phase-shifts. 

\subsubsection{The concept of $\bar{Z}$.}
The core charge $\bar{Z}$, and the core radius $r_c$ are important parameters in the pseudopotentials discussed here.
Some authors have expressed concern that $\bar{Z}$ is `not a well-defined
physical quantity', or that there is no quantum mechanical operator
whose mean value is $\bar{Z}$. One may also note that there is no quantum
mechanical operator whose mean value is the temperature $T$.
 Both $\bar{Z}$ and $T$ are Lagrange multipliers
used to define relational properties. Pseudopotentials, and 
many other calculational quantities have the advantage of being 
definable with respect to a convenient reference $H_0$, but this
does not mean any `arbitrariness' in the theory. The $H_0$ 
selected in pseudopotential theory enables us to reduce a many-electron
atomic problem to a few electron problem involving only $\bar{Z}$
electrons. 

A properly constructed $\bar{Z}$
should satisfy (i) the neutrality condition $n_e=\bar{Z}n_i$, 
where $n_i$ is the ion density. That is, $\bar{Z}$ is the Lagrange multiplier
whose value is chosen to ensures this neutrality condition. This is
readily generalized to a mixture of ions with different charges $z_j$ 
(ii) the  physical potential seen by a test charge in the
plasma should tend to $\bar{Z}/r$ for large $r$,
(iii) The $\bar{Z}$ should be consistent
 with the Friedel sum rule based on the phase-shifts from the ion with
a charge $\bar{Z}$ and a core radius $r_c$. 
 Finally, it has to be consistent with the set of
 bound states attached to the nucleus. These issues are discussed in
 Refs.~\cite{hyd0},\cite{elr0} and provide a stringent set of
 constraints in choosing $\bar{Z}$. We have found that these constraints could generally be well
 satisfied within finite-$T$ DFT calculations that use a sufficiently
large correlation sphere as the calculational volume.  It has to be
 recognized that a material system, e.g., an Aluminum WDM,
 could be a mixture of many ionization states, e.g.,  $Al^{z_j+}, z_j$
  =$\cdots-1,0,1,2,3 \cdots$  etc., where all ionization states are integers and 
 correspond to different descriptions of the core, with different values of
 $r_c$. The concentrations of each species would be such as to minimize the
 total free energy if it is an equilibrium system~\cite{elr0}.

\subsubsection{Yukawa potentials.}
The commonly used finite-$T$ Thomas-Fermi potentials,
also called Yukawa potentials, are
simple two-parameter forms. The core charge of the ions 
is $\bar{Z}$ and $r_c$ is set to zero (i.e., point ions).
The field electrons are characterized by a screening wavevector $K_Y$.
\begin{eqnarray}
W(r)_{ei}&=&-Z_Y/r\\
W(r)_{ii}&=&(Z^2_Y/r) \exp(-K_Y r)
\label{YukawaU.eqn}
\end{eqnarray}
We use atomic units with $|e|=\hbar=m_e=1$, and
the temperatures are in energy units.
 If the electron temperature
is $T_e$, the Thomas-Fermi electron-ion screening
 constant $K_Y$ is  given by
\begin{eqnarray}
\label{k-yukawa.eqn}
K^2_y &=&\frac{4}{\pi T_e}\int_0^\infty k^2dkn(k)\{1-n(k)\}\\
n(k)&=&1/\left[1+\exp\{(k^2/2-\mu)/T_e\}\right]
\end{eqnarray}
Here $\mu$ is the electron chemical potential at $T_e$ for the
electron density $n_e$ consistent with the ionization
$\bar{Z(T_i,Te)}$. In the simplest theory $\bar{Z}$ is determined
via finite-$T$ Thomas-Fermi theory. The structure factor
 consistent with this type of theory is:
\begin{equation}
S_y(k)=k^2/(k^2+K^2_Y)
\end{equation}

\subsubsection{DFT-based linear-response pseudo-potentials.}
The one-electron density is the
essential functional  that determines the equilibrium properties of a
quantum system. Hence the electron density $n(r)$ around a
nucleus of charge $Z$ placed in a uniform electron fluid at the bulk
density $n_e$ and temperature $T_e$ is computed. The computational
codes for such calculations have been available at least since 
Lieberman's {\it Inferno}  code~\cite{Liberman}, and extended recently
as in {\it Purgatorio}~\cite{Purgatorio}. Most such codes use a 
single ion in a Wigner-Seitz cell of radius $r_{ws}=\{3/(4\pi
\rho)\}^{1/3}$ as the computational volume when the ion density is
$\rho$. In contrast, we use a large correlation sphere
typical of the ion-correlations in the WDM. The single ion is
 surrounded by its self-consistent ion distribution $\rho(r)$ and the
corresponding electron distribution $n(r)$, calculated
self-consistently~\cite{hyd0}. A 
correlation sphere of radius $R_c$ of 5 to 10 times
$r_{ws}$ is used where $g_{ii}(R_c)\to 1$.
 In most cases one may use the simplified form due
to Perrot, where the ions are replaced by a uniform neutralizing
background except for a cavity around the nucleus~\cite{Pe-Be} to
mimic the ion-ion PDF. Further justification of this `neutral-
pseudo-atom' (NPA) model is given in ref.~\cite{eos95}

A weak local (i.e., $s$-wave) pseudopotential
$U_{ei}(q)$, where $q$ is the wave vector is calculated from the
free-electron density `pile-up' $n_f(r)$ obtained via the finite-$T$
Kohn-Sham equation. This $n_f(q)$ is the density
pile-up corrected  in NPA~\cite{eos95} for the cavity placed around
 the nucleus to simulate
the `cavity' of the $g_{ii}(r)$. Thus
\begin{equation}
\label{weak-eqn}
\Delta U_{ei}(q)=n_f(q)/\chi_{ee}(r_s, T_e, q)
\end{equation}
Here $\chi(r_s, T_e, q)$ is the electron linear-response function
 at the electron Wigner-Seitz radius  $r_s$, and temperature $T_e$.
A finite-$T$ local-field correction (LFC)  $G_q$ is also needed to
go beyond the simple random-phase approximation. That is,
\begin{eqnarray}
\label{chi-eqn}
\chi_{ee}(q)&=&\chi^0(q)/\{1-V_q(1-G_q)\chi^0(q)\} \\
\label{gq-eqn}
G_q&=&\{1-\gamma_0/\gamma\}(q/k_{TF})^2\\
V_q&=&4\pi/q^2,\; k_{TF}=(6\pi n/E_F)^{1/2}
\end{eqnarray}
Here $\chi^0(q)$ is the non-interacting
finite-$T$ Lindhard function. An electron effective mass $m^*$
(a `band mass') is used in evaluating the
Lindhard function. Thus, for Liquid Al near its melting point (m.pt., 0.081eV),
$m^*=0.998$ gives good agreement of the calculated $S(k)$
with experiment~\cite{Waseda}. At higher temperatures
 $m^*=1$. The LFC 
$G_q(T_e)$ is taken in the LDA (i.e., the $q\to 0$ form is used for all
$q$), and evaluated from the ratio of the non-interacting and
interacting finite-$T$ compressibilities $\gamma_0$ and $\gamma$ 
respectively.
 Alternatively, $G_q$ with full $q$-dispersion consistent
 with the $S(k)$ obtained from
the PDF $g_{ee}(r)$ may be used, as discussed in Ref.~\cite{prl1} .
 However, the formulation in terms of
the $q \to 0$  given in Eq.~\ref{gq-eqn} is quite accurate for
the systems studied here.

Instead of numerical tables obtained from Eq.~\ref{weak-eqn},
a simple fit to the $U_{ei}(q)$ may be used. The fits
used here are the Heine-Abarankov (HA)
forms~\cite{HA-ref}, or a slight generalization by
Dharma-wardana and Perrot~\cite{elr0}. The two elements Al
and Si chosen for this study only need a well-depth parameter $D$,
and $r_c$ which is the core radius. The electron effective mass
(i.e., band mass) $m^*$ 
enters into the response function.
\begin{eqnarray}
\label{pseudo-eqn}
U_{ei}(r)&=&-\bar{Z}D/r_c, \;\; r\le r_c \\ 
         &=&-\bar{Z}/r, \;\; r>r_c\\
U_{ei}(q)&=-&\bar{Z}V_qM(q)\\
M(q)&=&D\sin(qr_c)/qr_c+(1-D)\cos(qr_c)
\label{form-fac-eqn}
\end{eqnarray} 

Furthermore, it is found from test calculations that the scaled quantities 
$D/(\bar{Z}/r_{ws})$, $r_c/r_{ws}$ could be treated as constants for
changes in $r_{ws}$ if $\bar{Z}$ remains unchanged. This enables
 us to explore the behaviour of
these WDMs for compressions deviating significantly from unity, without having to do full DFT-calculations in each case. The use of a suitable
electron-effective mass $m^*$ bypasses  the complications of `non-local'
pseudopotentials and provide excellent results with just $s$-wave potentials.
Usually $m^*$ can be set to unity. It also may be used as a parameter fine tuned to get agreement with experimental $S(k)$ data, or in fitting a pseudopotential to $n_f(q)$ obtained from a DFT calculation, as in Eq.~\ref{weak-eqn}.

The NPA approach to simple local pseudopotentials is reliable if the
 resulting $U_{ei}(q)$ is such that
\begin{equation}
\label{pseud-good-eqn}
U_{ei}(q)/(-\bar{Z}V_q) \le 1.
\end{equation}

\subsection{Equilibrium pair potentials.}      
We present illustrative numerical results for Al and Si potentials
with $T_e=T_i=T$. 
The screened pair-potential $U_{ii}(q)$ is calculated from the
pseudopotential via: 
\begin{equation} \label{pairpot-eqn}
U_{ii}(q)=\bar{Z}^2(T)V_q+|U_{ei}(q)|^2\chi_{ee}(q, r_s, T)
\end{equation} 
Further improvement of these formulae to incorporate
core-polarization effects etc., via DFT are discussed in the appendix of
Ref.~\cite{eos95}.

Potentials for calculations with C, Si, and Ge in the equilibrium
WDM regime have
been discussed by Dharma-wardana and Perrot in Ref.~\cite{csige90},
and for Al  for $T>5$  eV in Ref.~\cite{elr0}. The normal-density
liquid-metal Al pseudopotential used here is the potential discussed
in Ref.~\cite{Dharma-Aers83}. This $U_{ei}(q)$ generates the DFT $n(q)$
under linear response and recovers experimental $S(k)$ data quite well.
 
\begin{figure}
\includegraphics*[width=8.5 cm, height=11.0 cm]{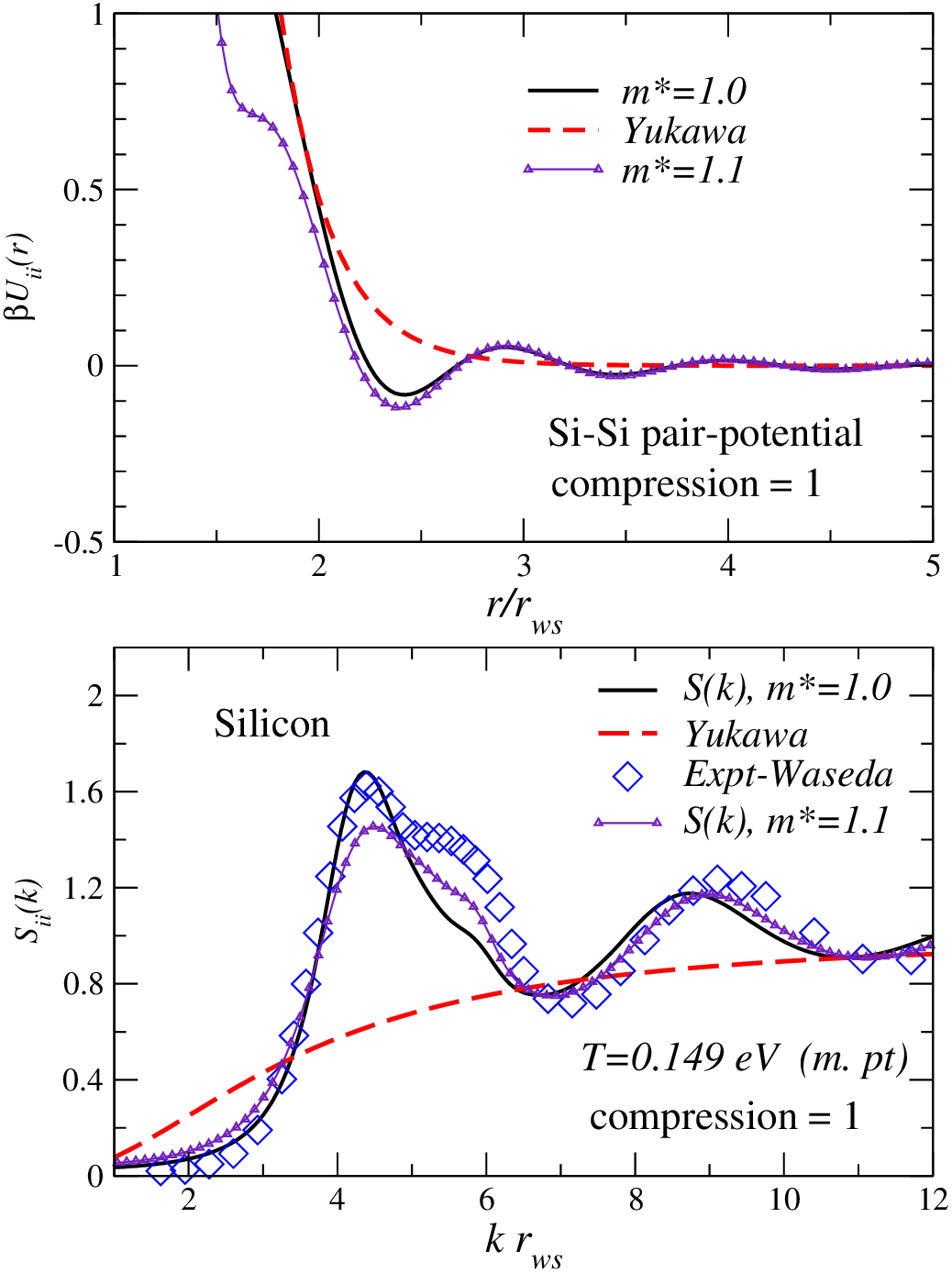}
 \caption
{(Online color).(top) The Si-Si pair-potentials $\beta U_{ii}(r)$,
$\beta=1/T$, at the melting
 point ($T_i=T_e=T= 0.15$ eV) at unit compression ($r_{ws}$=3.07au), using
 an empty-core pseudopotential with core radius $r_c/r_{ws}$ =0.3084
 are displayed for electron effective masses $m^*=1$, and 1.1. The
 Yukawa potential~\ref{YukawaU.eqn} is also displayed. (Bottom panel)
 The corresponding ion-ion structure factors $S_{ii}(k)$, as well as the
 experimental data of Waseda~\cite{Waseda} are shown. The additional
 structure near the Kohn anomaly (~2$k_F$) beyond the main-peak is 
 very sensitive to the local band mass $m^*$} 
\label{SiUii-fig}
\end{figure}

Pair potentials of other elements and at most compressions can
 be evaluated as in
Ref.~\cite{Pe-Be}.  Numerical results for $\beta U_{ii}(r)$, $\beta=1/T$
are displayed in
  Fig.~\ref{SiUii-fig}.  They are are
particularly interesting in view of the
subsidiary $2k_F$ peak in $S(k)$. Solid silicon has a rather open
structure (diamond-like) which collapses under melting into a
high-density metallic fluid. Car-Parinello simulations, many-atom
Stillinger-Weber type potentials (requiring many empirical parameters)
etc., need to be highly elaborate to capture the subsidiary
structure associated with the electron mass $m^*$ and the 2$k_F$
scattering processes. In fact, attempts were made in some of these
 early MD studies to explain this `hump' in the Si-$S(k)$ in terms
 of covalent bonds between Si atoms. In reality,  molten Si is an
excellent metal and the `hump' is a metallic property associated
with $m^*(k)$. In our view, two-center bonding effects do not
arise unles electron
densities fall to low values (e.g., 1/64 of the normal
electron density for Al), and the temperatures become sufficiently low.

WDM systems generated from laser pulsing or shock compression
usually starts from a solid-density. Thus the pseudopotentials
and pair-potentials should successfully recover the strongly-correlated
low-temperature regime near the melting point (m.pt), as well as the
higher-$T$ and $\kappa$ regimes. Our Al-Al pair potential at the m.pt
is shown in the inset to Fig.~\ref{AlUii-fig}, together with the DRT
potential (based on a non-local pseudopotential) 
which is known to recover the properties of solid and liquid Al at 
low-temperatures~\cite{DRT}.

It is seen that the Yukawa potentials approximately reproduces the
envelope of the pair-potentials, while failing to reproduce the
Friedel oscillations and energy minima. Similarly, the Yukawa
structure factor $S_y(k)$ does not have oscillatory features.

\subsubsection{Pair-potentials for mixtures.}
\label{mixtures-sect}
Many WDM systems involve mixtures of ions. Thus plastic converts to
a compressed WDM containing mostly a mixture of H, C, O ions. Even
an Aluminum plasma could contain several stages of ionization and
they have to be treated as a mixture. The detailed treatment of such
systems as mixtures of average atoms, their chemical species-dependent
 potentials, embedding
energies etc., as well as pair-potentials were discussed in
Ref.~\cite{eos95}. The pair potential involving two types of ions
carrying mean charges $\bar{Z}_a, \bar{Z}_b$ could be written as 
\begin{equation}
U_{a,b}(q)=V_q\bar{Z}_a\bar{Z}_b+|U_{ea}(q)U_{eb}(q)|\chi_{ee}(q)
\end{equation}
Such cross-potentials would be inputs to the MHNC equations, or to
MD simulations, for generating the corresponding PDFs.

\subsection{Non-equilibrium pair-interactions.}
\label{noneq-pair-subsec}
In the following we examine two-temperature quasi-equilibrium
systems rather than full non-equilibria. Such
systems occur during times $t$ such that $\tau_{ee}<\tau_{ii}< t <\tau_{ei}$,
after the energy is dumped in either the ion subsystem (mechanical
shock) or in the electron subsystem (by a laser pulse).

\subsubsection{Laser-pulse generated quasi-equilibria.}
\label{laser-pulse-sub}
When a metal foil at the bath-temperature $T_b$ is subject to a
short-pulse laser, the electron temperature rises rapidly to some
$T_e$ within   femto-second times scales $\tau_{ee}$, by
equilibration via electron-electron interactions. The transfer of
energy to the ion subsystem, arising from electron-ion interactions 
is slow, and hence the ion temperature $T_i$  remains essentially
locked to $T_b$, while the electrons reach $T_e$. Hence, processes
occurring within the  electron-ion temperature relaxation timescale
$\tau_{ei}> t> \tau_{ii}$  may be probed to provide information about
 quasi-equilibrium systems. The probes are
optical pulses sampling a volume related to a space-time average
over the pulse time and the optical depth of the material.  Hence
the `experimental results' reported should be regarded as already
implicitly containing some sort of interpretational model.

 A proper
description of such systems needs  pair-potentials
$U_{ii}(r, T_e, T_i)$ where $T_e\ne T_i$. 
The procedure described for the equilibrium system can be simply
generalized for quasi-equilibria. In
laser-heated systems, the pseudopotential is that of the initial
state $T=T_i=T_e$, and this remains unchanged under changes
 of $T_e$ as long as
$\bar{Z}$ remains unchanged.  
Unlike in the equilibrium case, the bare ion-ion pair potential in
a laser-heated metal is screened by the (hot) electrons at $T_e$.
Hence,
\begin{equation}
U_{ii}(q, T_i, T_e)=\{\bar{Z}(T_i)\}^2V_q+
|U_{ei}(q,T_i)|^2\chi_{ee}(q, r_s, T_e)
\end{equation}
The ion-ion
structure factors $S(k)$ is that of the initial state
at $T=T_i=T_e$ for time scales shorter than $\tau_{ei}$.  

\begin{figure}
\includegraphics*[width=8.5 cm, height=11.0 cm]{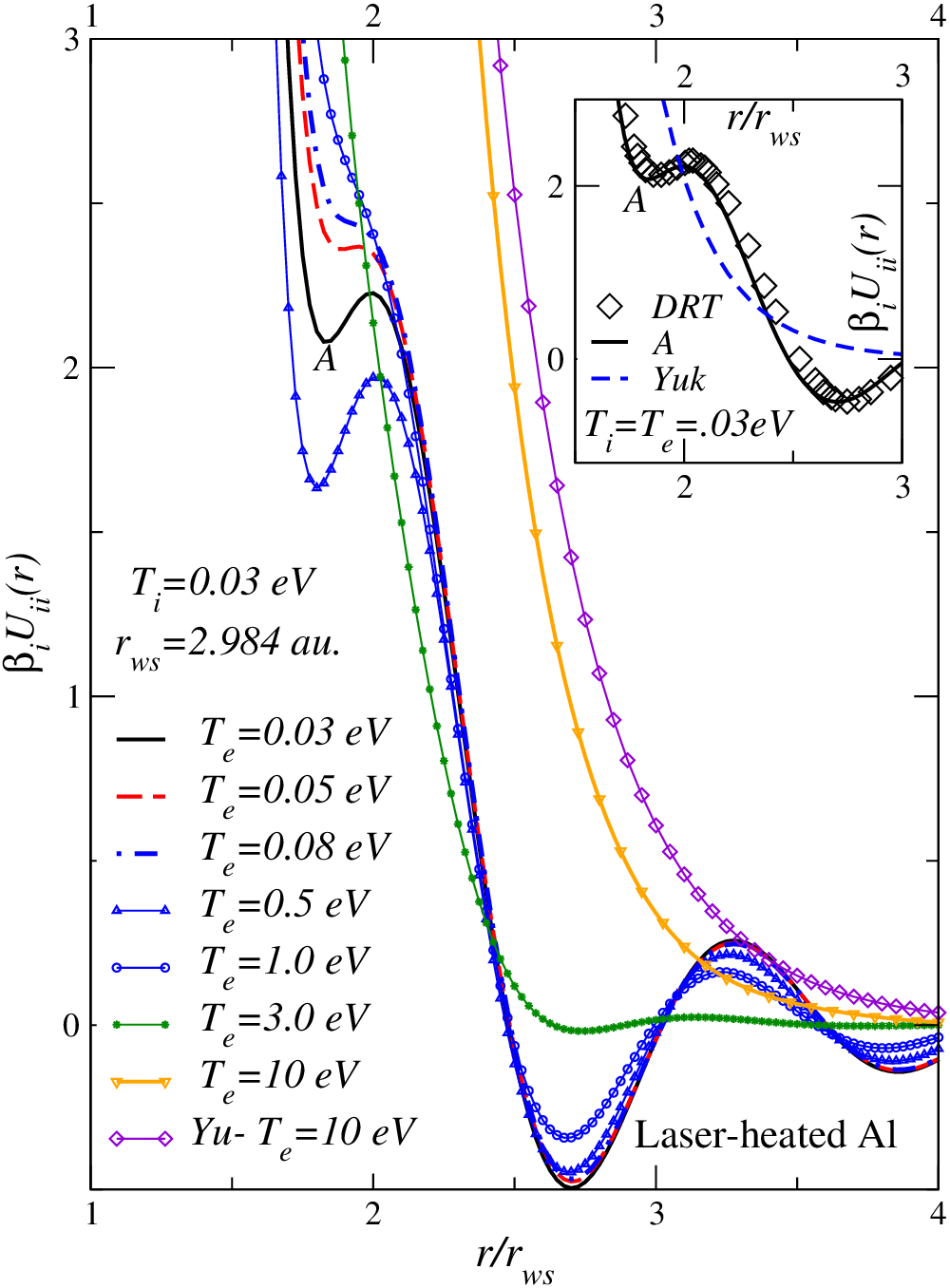}
 \caption
{(Online color). (Inset) The Al-Al pair-potential, $\beta_iU_{ii}(r)$,
$\beta_i=1/T_i$,
  at room temperature
 ($T_e=T_i=T$ = 0.03 eV) and normal density, labeled {\it A} is derived from a
two-parameter local pseudopotential and a band mass $m*=1.02$. It is 
compared with a well established non-local potential ~\cite{DRT} that
accurately reproduces structural and phonon data of Aluminum. The
 Yukawa potential, Eq. ~\ref{YukawaU.eqn}, at $T=0.03$ eV 
is also displayed.
 The main panel  displays the evolution of the potential
  while $T_i$ is held fixed
 at 0.03 eV, and $T_e$ is increased. The potential becomes repulsive
 from $T_e=$0.03 eV to 0.08 eV. For $T_e>0.08$ eV it deepens
  and becomes attractive, e.g., at $T_e=0.5$ eV, before it finally
 becomes repulsive. The Yukawa potential at $T_e=10$ eV is also
 shown} 
\label{AlUii-fig}
\end{figure}

The main part of Fig.~\ref{AlUii-fig} shows
the evolution of the potentials under laser-pulsed heating
 where $T_e$ increases while the ion subsystem remains
 intact for time scales
shorter than $\tau_{ei}$. The
first minimum near $r/r_{ws}\sim 1.85$ (marked {\it A}) weakens at first,
and then deepens for $T_e>\sim 0.08$ eV (with $T_i$ fixed at 0.03 eV). This is
clearly evident for the case $T_e=0.5$ eV in the figure. This may
 be considered a microscopic signature of `phonon hardening' observed
  in some experiments~\cite{Ernstorfer09}. 

 The highest
phonon frequency $\omega_{ph}$ supported by the Al-lattice at 90K is
$\sim 6\times 10^{13}$ s$^{-1}$, while this decreases  rapidly at
higher temperatures, and becomes purely imaginary if the lattice
melts.
The concept of a `phonon' may not become meaningful if
$1/\omega_{ph}$ begins to exceed $\tau_{ei}$. In fact, the
two-temperature pair-potential itself requires updating at
 each time step.   

When the electron temperature rises above $\sim 1$ eV,
 the potentials (see Fig.~\ref{AlUii-fig}) become repulsive 
and the resulting compression
  in the electron subsystem increases rapidly,
leading to a  `Coulomb explosion'. We have assumed that
$T_e$ is not sufficiently high to open up an ionization channel that
would add to the energy relaxation between cold ions and hot
electrons. Such processes may be addressed as in Ref.~\cite{mmll}.

\subsubsection{Shock generated quasi-equilibria}
\label{shock-sub}
The time evolutions of $T$ and $\kappa$
in shock-compressed systems are quite different to those of
laser-heated samples. The Hugoniot is the locus of states that can be
reached by a single shock wave, providing a relation connecting the
volume, internal energy and the compression. Well-developed
mathematical and numerical techniques exist for shock
studies, if the equation of state (EOS) is 
available~\cite{Swift07}. The Hugoniot
prevails only after many time steps greater than $\tau_{ei}$.
 We are interested in shorter quasi-equilibrium
time scales. Experimentally, the temperature evolution of the ion
 subsystem is very difficult to measure, and the temperature deduced
  from the thermal
emissivity of the shocked state is for the state following 
hydrodynamic interactions between the sample and the observation
window.

Thus material properties of the shocked state is
determined from optical measurements combined with velocities of
shock fronts and particles.  
The `sample
region' (seen by the optical probe) is the layer of material within
approximately one optical depth behind the shock front (assuming
that we are observing a shock wave in-flight inside the sample with
no release wave).  For a typical optical depth of $\sim$ 5 nm, and
shock speeds of $\sim 10^4$ m/s, the transit time $\tau_{sh}$ of
 the shock front
 through this layer is $\sim$ 0.5 ps~\cite{AndrewNg12}.
Since $\tau_{ee}<<\tau_{sh}$, there is a meaningful $T_e$ in the
sample region. In any case no energy has been dumped into the
electron subsystem for $t << \tau_{ei}$. Since the phonon frequency
of typical solids is of the order of several THz (i.e.,
$\tau_{ii}<\tau_{sh}$), one may assume that the ion subsystem can be
characterized by a temperature $T_i$.  
This would
lead to a single two-$T$ quasi-equilibrium in the sample volume,
 or possibly a  gradient of such quasi-equilibria in $T_e$
  and $T_i$ along the path
behind the shock front. In either case, we need two-temperature
pair-potentials along a non-equilibrium `Hugoniot'.

\begin{figure}
\includegraphics*[width=8.5 cm, height=12.0 cm]{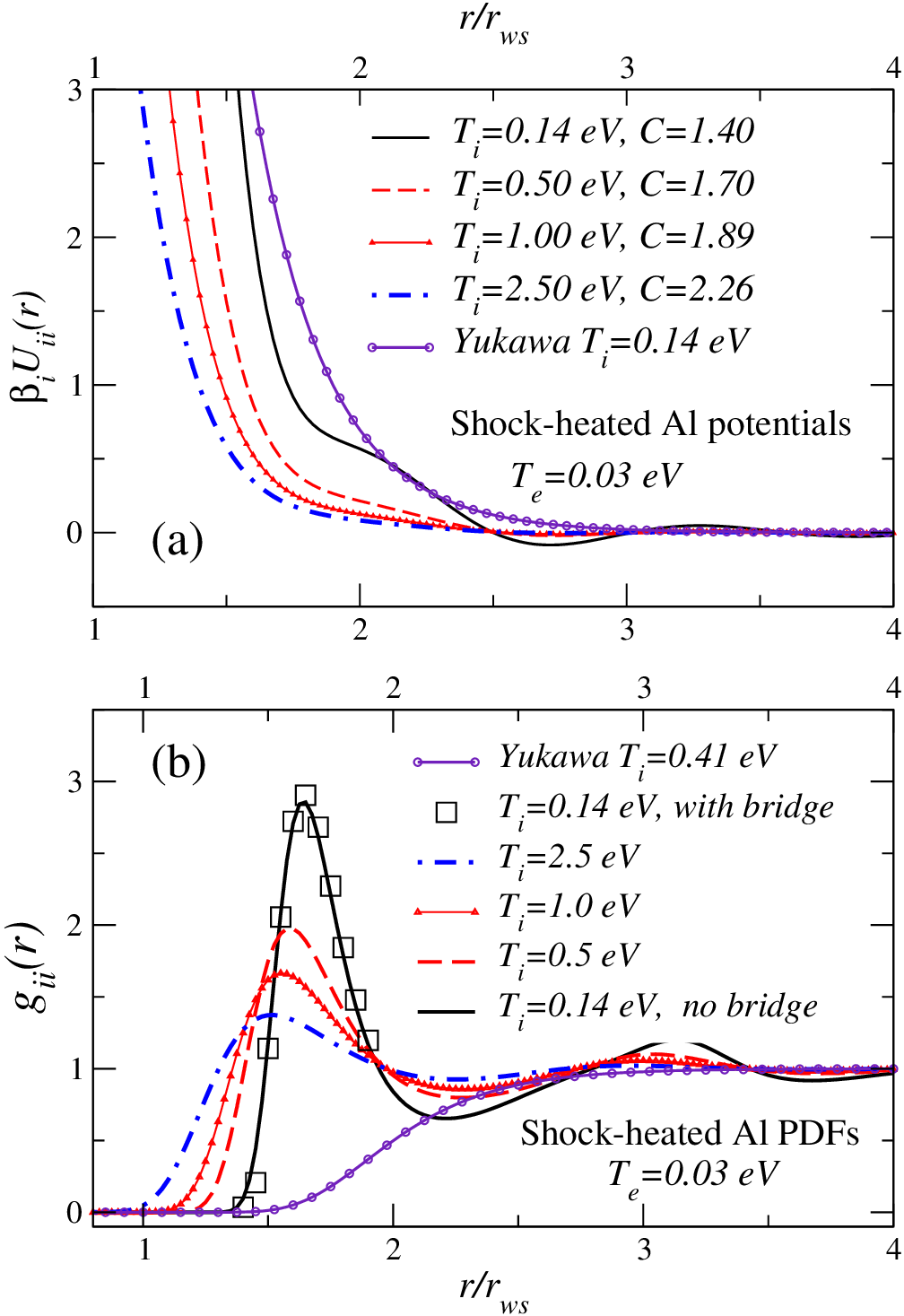}
\caption
{(a)The pair-potentials $\beta_i U_{ii}(r), \beta_i=1/T_i$
for a shock-heated Aluminum WDM with an
initial temperature of $T_e=0.03$ eV, with the ion-subsystem
temperature increasing to 2.5 eV. (b) The corresponding ion-ion PDFs
$g_{ii}(r)$  where $T_e$ remains fixed at 0.03 eV,  while $T_i$
increases to 2.5 eV as displayed. Calculations with bridge (MHNC)
 and without bridge
corrections (HNC) are given for $T_i=0.14$ eV, showing that bridge
corrections are negligible. The other PDFs are HNC solutions. The
Yukawa PDF is the Boltzmann form.}
\label{Ushock-fig} 
\end{figure}

We model this system as follows. Here the usual EOS is a good approximation
 to the quasi-EOS
mainly because the energy in the electron system is small compared to that of
the ions. Electrons remain more or less degenerate even at
$T_i$=10 eV, when the compression  reaches $\sim 3$. If $\kappa$ remains
within a region where $\bar{Z}$ is unchanged, the bound electrons
remain intact. However, in general the value of $\bar{Z}$
appropriate to the compression must be used and energy-relaxation by
ionization need to be considered~\cite{mmll}. 
 The compressions are
calculated from Al and Si Hugoniots where we may neglect the small differences
in the contribution to the internal energy from the electron subsystem
at equilibrium or at quasi-equilibrium. The Al Hugoniot is based on our
neutral-pseudoatom calculations~\cite{eosbenage}.
 The Si Hugoniot~\cite{Stern} is the {\it L140}
data based on the Quotidian-EOS model~\cite{L140}.

The onset of compression makes the potential
 take a Yukawa form except at very low-energy deposition.
 This pair-potential does not support
phonons or `phonon hardening' at  the accessible short-times scales,
or at longer time scales. In this problem the ion-electron time
scales $\tau_{ei}$ are lengthened due to the formation of
ion-acoustic coupled modes involving both electrons and ions (hence
the life times of the two-$T$ quasi-equilibria become
longer).
\begin{table}
\caption{The temperature $T_i$ (eV) and compression $\kappa$, 
for shock-compressed aluminum and Si, for an initial normal density and
temperatures $T_e$ = $T_i$ = 0.03 eV. 
 The ionization $\bar{Z}$ remains essentially
 at 3 for Al, and 4 for Si. The electron temperature $T_e$
remains at the initial temperature for time scales less
 than $\sim \tau_{ei}$.
}
\begin{ruledtabular}
\begin{tabular}{cccccccc}
$T_i (eV)$        &0.14   & 0.25  & 0.50  & 1.00 & 2.50  & 5.00  \\
\hline\\
$\kappa$-Al  &1.4    &1.57   &1.70   &1.89  &2.26   &2.59\\ 

$\kappa$-Si  &1.57   &1.70   &1.87   &2.14  & 2.65  &3.07\\
\end{tabular}
\end{ruledtabular}
\label{hug-tab}
\end{table} 

 Figure~\ref{Ushock-fig} shows the pair-potentials
(upper panel) and the corresponding two-temperature 
pair-distribution functions
obtained by this procedure for Al. No bridge corrections have been included
in the $g(r)$ shown in Fig.~\ref{Ushock-fig} except for $T_i=0.14$ eV,
 $T_e$ = 0.03  eV,
where it is shown that their effect is
negligible. The Yukawa PDF is taken as the Boltzmann form 
$\exp\{-W_Y(r)/T_i\}$.

One may wonder whether potentials beyond 
binary-interactions would be necessary in a classical representation
of a many-particle system. Attempts to simulate classical fluids
with model potentials (e.g., Lennard-Jones) show that binary
interactions are often inadequate. However, since the potentials
were made out of Coulomb interactions in a systematic manner using
DFT densities,  we
believe that such effects are not significant, and probably get
included in the problem of modeling the bridge functions. However,
this needs further study, as it could be
important in low-density, low temperature systems, and near the critical
point~\cite{DW-Aers-Kr, Rosen-Kahl97}.

\section{two-temperature transport coefficients.}
\label{two-T-transport}
The transport coefficients that are of common interest are
associated with weak applied gradients in the electric potential,
temperature etc., so that linear transport coefficients like the
electrical conductivity $\sigma_e$ and heat conductivity $\sigma_h$
could be defined. However, if the non-equilibrium system is such
that the subsystem Hamiltonians $H_e$, $H_i$ are invariant only for
time scales $\tau_{ee}$, $\tau_{ii}$, then the $\omega \to 0$ limit
of the transport coefficients, i.e., their static values are not
defined in any strict sense. Thus extrapolations of experimental data
to the $\omega \to 0$ have to be carefully examined to understand
their physical content. We refer to any processes with a time scale
$\tau$ such that $\tau_{ii} <\tau < \tau_{ei}$ as quasi-static. 

Thus these experiments need to be planned, keeping in mind the available
lifetimes for experimentation. Evaluating energy-relaxation lifetimes
 $\tau_{ei}$ from the e-i pseudopotentials and dynamic response functions
is itself an extremely demanding task~\cite{elr0,mmll} 
but needed in WDM studies. We note that if the
Al-pseudopotential discussed here is used at unit compression, $T_i$ = 0.1 eV, 
$T_e =$ 1.0 eV, then $\tau_{ei}$ is $\sim 6.02\times10^6$
electron-plasma oscillations long, while at $T_i$ = 0.1, $T_e$ = 5 eV this
reduces to $\sim 0.72$ of the value at $T_e=1$ eV. These numbers are based
 on our formula for the energy-relaxation rate evaluated via the
{\it f}-sum rule~\cite{mmll}, without including coupled-mode effects.

Here we examine the theory of the dynamic two-temperature conductivity
as a typical example. The objective is to extract a
 frequency-independent (i.e.,
quasi-static) component, and present numerical calculations for the
quasi-static part of the resistivity using the pseudopotentials and PDFs
developed in the previous sections. This analysis puts a different
perspective on the physical content of the  two-temperature Ziman
 formulae that have been used in the past for these systems.
 The extent of its physical validity
 has to be established by experiment.

\begin{figure}
 \includegraphics*[width=8 cm, height=11 cm]{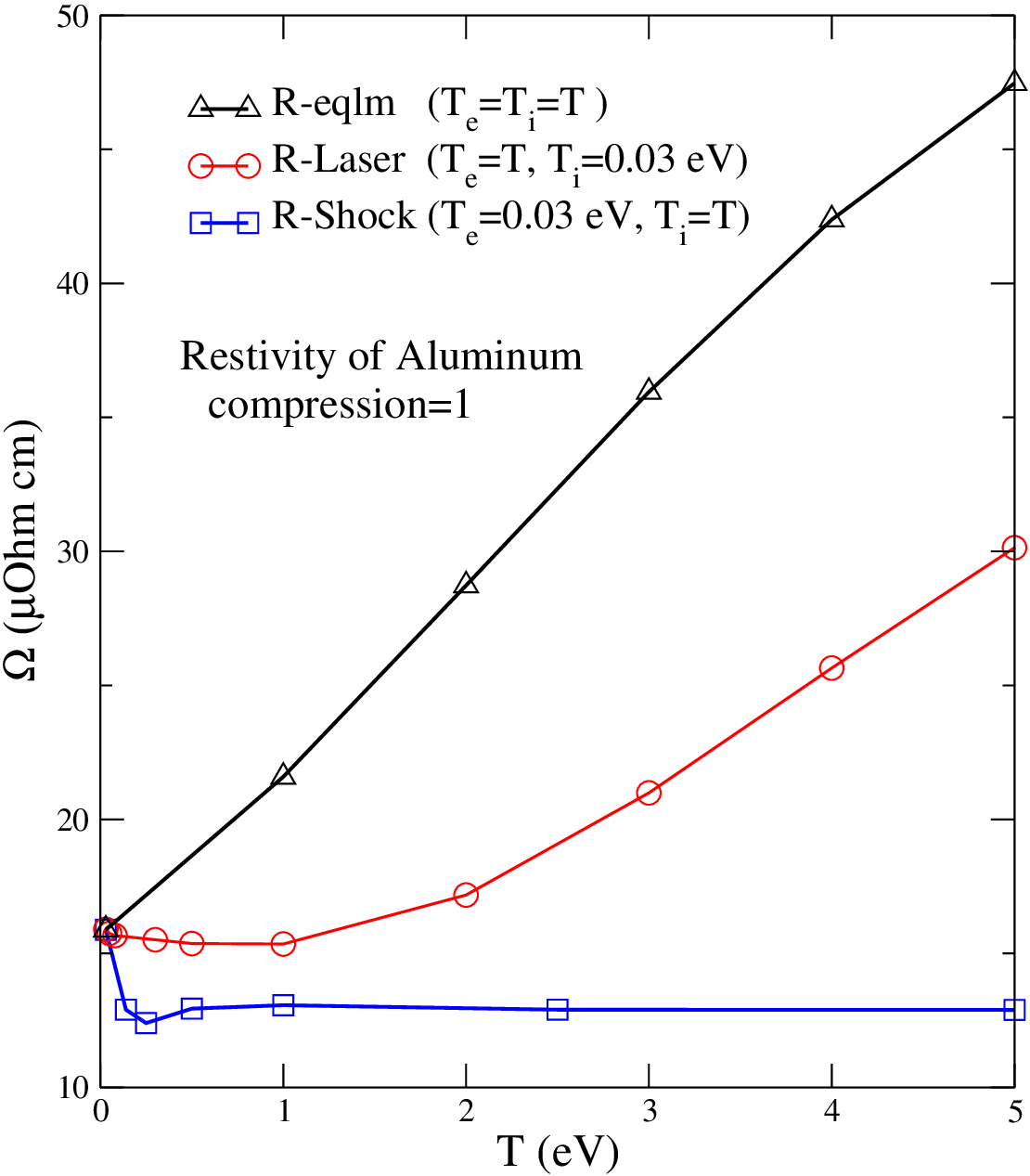}
\caption
{(Color online). Triangles: the
equilibrium static resistivity of Aluminum at normal compression when
$T=T_e=T_i$ increases from the melting point. Circles: the evolution
 of the quasi-static component of
the dynamic resistivity (Eq.~\ref{two-T-Ziman-eqn}) of WDM-Aluminum
at the melting density, with the ion-temperature $T_i$ held at 0.03 eV
while the electron temperature $T_e$ is
increased, as in laser-heated WDMs. Squares:  the corresponding
quasi-static resistivity for a shock-heated Al-WDM where $T_e$ remains
fixed at 0.03 eV, while $T_i$ increases  to 5 eV with compression,
where $\kappa$ increases from 1.0 to $\sim 2.6$ . }
\label{R-Al.fig}  
\end{figure}

The response time of the system to the probe has to be smaller than
the system relaxation time $\tau_{ei}$, and long enough for the
particle distributions to become stationary under the probe field. 
If a quasi-static electric field $\vec{E}$ is applied to an
electron distribution $n(k)$, where $k$ is a wave vector, then 
it is the displaced {\it stationary} distribution
$n(k,E)=n(k)+\delta(k,\vec{E})$ that determines the electrical
conductivity of the system. The probe speed has to be consistent with the
formation of a steady-state displaced-electron-distribution $n(k,\vec{E})$
in the electron system. The probe frequency cannot be slower
 than  $1/\tau_{ei}$. 
Hence quantities obtained by  extrapolating $\omega \to 0$
 may not have a physical meaning. 

\begin{figure}
\includegraphics*[width=8 cm, height=11 cm]{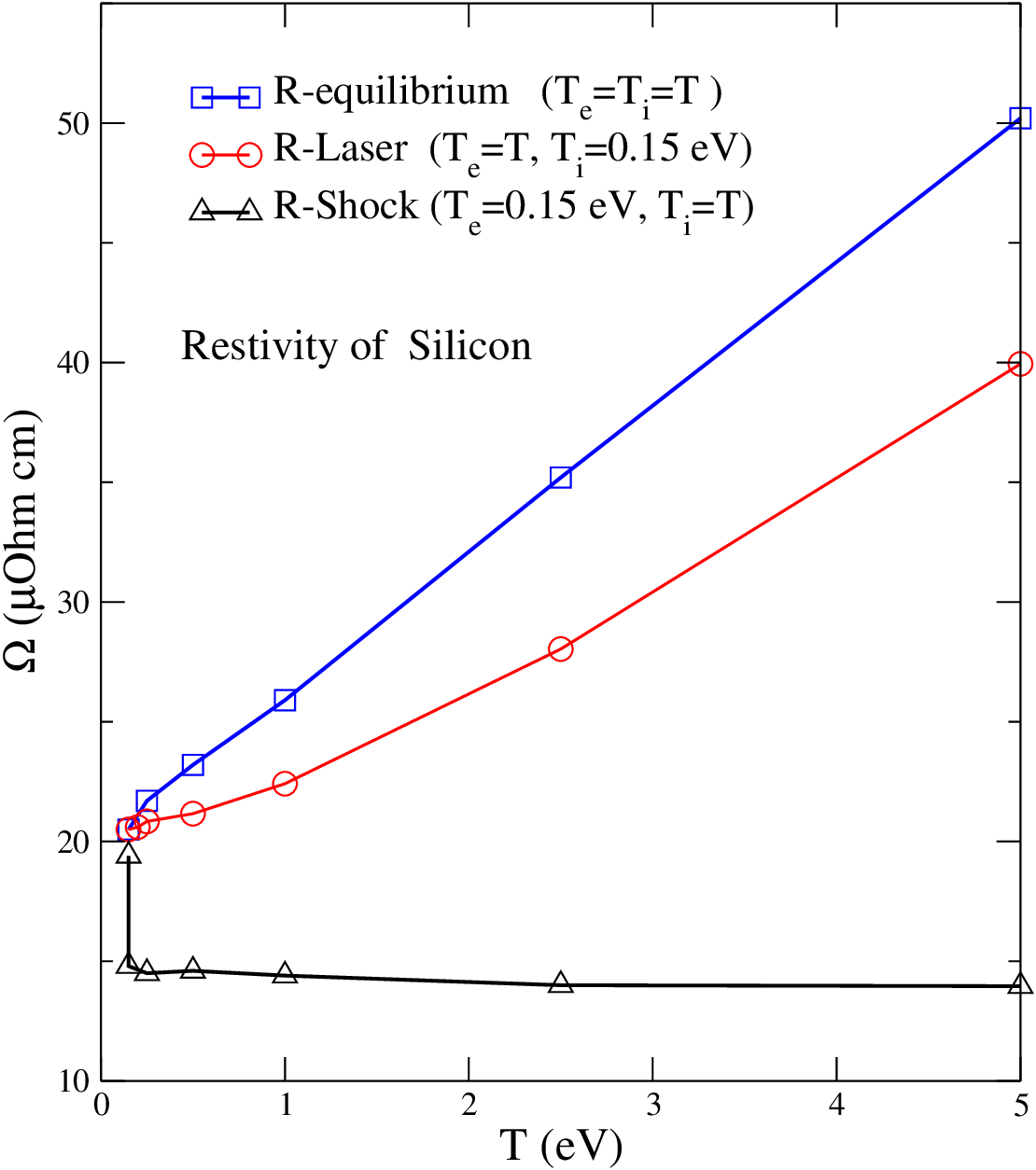}
\caption
{(Color online). Triangles: the
equilibrium static resistivity of Si at normal compression when
$T=T_e=T_i$ increases from near the melting point. 
Circles: the evolution
 of the quasi-static component of
the dynamic resistivity  (Eq.~\ref{two-T-Ziman-eqn}) of WDM-Si
at the melting density, with the ion-temperature $T_i$ held at 0.15 eV
while the electron temperature $T_e$ is
increased, as in laser-heated WDMs. Squares:  the corresponding
quasi-static resistivity for a shock-heated Si-WDM where $T_e$ remains
fixed at 0.15eV, while $T_i$ increases  to 5 eV with compression,
where $\kappa$ increases from 1.0 to $\sim 3$.} 
\label{R-Si.fig} 
\end{figure}

The measured conductivity can be associated with a
scattering time $\tau(\omega)$ by writing
\begin{equation}
\label{drudetau-eq}
\sigma(\omega)=\frac{ne^2 \tau(\omega)}{m_e\{1-i\omega\tau(\omega)\}}.
\end{equation}
 Here we have restored the electron charge and mass
although we use atomic units. The conductivity $\sigma(\omega)$ can
be expressed in terms of the Fourier component at the frequency
$\omega$ of the current-current correlation function $<<J(t),
J(t')>>$. A standard evaluation, assuming that the effect of the
total system is a superposition of the effect of individual
scatterers can be carried out,  as in the equilibrium
case~\cite{Mahantxt}, but using the Keldysh contour for the
non-equilibrium case. Here we  keep in mind that the subsystem
temperatures are $T_e$ and $T_i$, giving the result:
\begin{eqnarray}
\label{dynamic-tau-eq}
\frac{1}{\tau(\omega)}&=&-\frac{1}{(\omega^e_p m_e)^2\omega}
 \int_0^\infty\frac{q^2 dq }{(2\pi)^3}|U_{ei}(q, r_s, T_e)|^2q^2_z\nonumber\\
 &&\times \int_0^\infty \frac{d\nu}{2\pi}Im\{\chi_{ii}(q,\nu, r_{ws}, T_i)\}
\Delta N(\nu,\omega)\nonumber \\
 &&\times Im\{1+V_q\chi_{ee}(q,\omega+\nu, r_s, T_e)\}\\
\Delta N(\nu,\omega)&=& [N((\nu+\omega)/T_e)-N(\nu/T_i)]
\end{eqnarray}
%
%
The last line contains Bose occupation factors
$N\{(\nu+\omega)/T_e\}$ and $N(\nu/T_i)$. Here $U_{ei}(q,r_s,T_e)$
is the pseudopotential presented in Eqs.~\ref{weak-eqn}-\ref{pseudo-eqn}. The
frequency-dependent ion response function and the electron response
function also appear. The latter was discussed in
Eq.~\ref{chi-eqn}.  The ion-response function can be constructed in
a similar manner, or obtained from an MD simulation~\cite{Jacucci}. 

This expression for the dynamic conductivity or $\tau(\omega)$ differs
significantly from the Green-Kubo (GK) form used by a number of
authors~\cite{mazevet} since both electron-electron and ion-ion
many-body effects are treated dynamically in Eq.~\ref{dynamic-tau-eq}.
 In particular, in the
currently available GK implementations, the Green-Kubo
formula is averaged within an ensemble of {\it static}
configurations generated by  ion
molecular dynamics; the electrons are quenched to a Born-Oppenheimer
surface to evaluate one-electron Kohn-Sham eigenvalues which are
used in the GK formula. The latter actually calls for eigenvalues of
the Dyson equation. The 'quenching process'
destroys dynamical information;  the ions and electrons `do not know' each 
other's temperatures.

We extract a meaningful `static' component of the quasi-equilibrium dynamic
resistivity as this has been the object of recent experiments, e.g.,
in Ref.~\cite{NgAu}. The physically meaningful lowest value
$\omega_L$ of the probe frequency $\omega$ is $\sim 1/\tau_{ei}$.
Hence, when $T_e$ is not strongly  different from $T_i$ we may use 
$\omega=\omega_L$ in a Taylor expansion to write:
\begin{equation}
\label{occup-taylor-eqn}
\Delta N(\nu,\omega)=N(\nu/T_e)-N(\nu/T_i)+
\omega \frac{\partial N(\nu/T_e)}{\partial \nu} +\cdots
\end{equation}
The leading term in Eq.~\ref{occup-taylor-eqn} is zero in the
equilibrium case $T_e=T_i$. It is seen that the
first-derivative term gives the Ziman resistivity in the 
equilibrium case, where an $\omega \to 0$ limit exists. When $T_e \ne
T_i$, the first term in $\Delta N$ is non-zero
and leads to a divergence unless $\omega$ remains finite.
Thus the concept of a static conductivity is non-physical unless
this `zeroth-order' correction could be deemed negligible.

Replacing $\omega$ by its least allowed value $\omega_L\sim
1/\tau_{ei}$ in the above equation and in the pre-factor in
Eq.~\ref{dynamic-tau-eq}, we obtain a frequency-independent
first-order contribution. This can be further reduced to a
two-temperature Ziman resistivity of the form:
\begin{eqnarray}
\label{two-T-Ziman-eqn}
R(T_e,T_i)&=&\frac{\rho(T_i)}{3\pi n(T_e)}\int_0^\infty d\epsilon
 \frac{d f(\epsilon)}{d \epsilon}
\int_0^{2\surd{\epsilon}}S_{ii}(q, r_{ws}, T_i)\nonumber \\
 && \times\left|\frac{U_{ei}(q,r_s,T_e)}{\varepsilon(q,r_s,T_e)}\right|^2q^3dq\\
\varepsilon(q,r_s,T_e)&&=1+V_q\chi_{ee}(q,r_s, T_e)
\end{eqnarray}
The ion-ion structure factor $S_{ii}(q)$ is obtained from the
ion-ion PDF via the Fourier transform of $g_{ii}(r)-1$. We emphasize
that this quasi-static component of the dynamic resistivity is {\it not}
 the $\omega \to 0$ limit of the dynamic resistivity.  
  
The numerically inconvenient integration over the derivative of the
Fermi function in Eq.~\ref{two-T-Ziman-eqn} can be side-stepped by
 using the form of the Ziman
formula given by Perrot and Dharma-wardana, viz., as Eq.(31)
in Ref.~\cite{eos95}.

Whether the static part of the dynamic conductivity extracted in the
above manner would be the quantity obtained by extrapolation of the
experimental results is unknown. In Figs.~\ref{R-Al.fig} and
~\ref{R-Si.fig} we
display the variation of $R(T_i,T_e)$ for normal density Aluminum and Silicon at
the melting point under laser heating, contrasted with 
shock heating where the variable is $T_i$. The compression remains at unity
in the laser-heating and `equilibrium' curves, while the compression rises
in the shock-heating experiment. Similar results for
Si are given in Fig.~\ref{R-Si.fig}

It should be noted that all resistivity calculations, even for
equilibrium systems, may differ significantly from one another depending on 
the specific (Boltzmann conductivity or the Ziman resistivity) formula
that is used. They themselves may differ significantly, depending whether a
$T$-matrix or a pseudopotential is used~\cite{PDW-Thermophys}.
If a $T$-matrix is used, corresponding modified densities of states should be
used (here we have used the simple pseudopotential only).
However, the  trends shown in Fig.~\ref{R-Al.fig} and~\ref{R-Si.fig}
clearly distinguish between equilibrium, laser-heated and shock-heated
plasmas.

\section{Discussion}
\label{disc-sec}
We have examined the construction of effective
potentials for electron-ion and ion-ion interactions that take
account of the ambient material conditions in warm-correlated
matter, since standard pseudopotentials are ill-adapted to
finite-$T$ problems. 

The effective potentials can be used in classical MD simulations or
modified hyper-netted-chain equations to obtain pair-distribution
functions. The PDFs or
calculated structure factors have been used,
together with the pseudopotentials
to obtain free energies, Hugoniots, energy relaxation and
transport coefficients, even in quasi-equilibrium situations,
with minimal computational effort. The accuracy of the methods could be
tested against liquid-metal WDM data. The $T_e\ne T_i$
results presented here have shed light on
issues like phonon hardening, and quasi-static resistivities. 

An alternative approach is to use statistical potentials generated
via  Slater sums~\cite{cimarron, Ebeling2006}. However, as
far as we know, those methods
have not been tested against experimental liquid-metal $S(k)$ and transport
 date. The present
 approach based on the neutral-pseudo-atom DFT density is 
conceptually and computationally simple. The
non-linear response of the electron system to a nucleus is
treated by DFT, and these non-linear features are included
in Eq.~\ref{weak-eqn} where finally a {\it linear-response}
pseudopotential is constructed. The weak 
pseudopotentials presented here enables us to write down the
pair-potential directly,  where as the pseudopotentials provided
with standard codes need a further solution of the Schr\"{o}dinger
two-ion problem to extract a pair potential. That approach is indeed
necessary when the condition given in Eq.~\ref{pseud-good-eqn} is
not satisfied. However, in the cases considered in this paper, the
full Schr\"{o}dinger two-ion problem, as well as the multi-ion
problem (e.g., as realized in Car-Parinello DFT-MD calculations on
C, Si, Ge molten fluids~\cite{csige90}, and on Al-WDM~\cite{silv1})
yield results in good agreement with those obtained from our single-center
methods.

The ranges of validity of the methods discussed here are
 those of (i) the underlying DFT code used to construct the
self-consistent charge density $n(r)$ around {\it a single} nucleus. This
is not valid if chemical bond-formation is possible, e.g., at
low free-electron densities and low temperatures. (ii) the assumption that a
linear-response pseudopotential exists, (iii) and that the relevant time
scales are satisfied. High-$Z$ materials like Au, W, pose special
difficulties by these (or other) methods. The theory 
of equilibrium-Au WDM by these methods is given in Ref~\cite{lvm}.
 The main improvement needed is  a
relativistic-DFT calculation~\cite{grant}. Once the relativistic
 charge distribution $n(r)$ is obtained, a weak
pseudopotential may not exist. Then a two-center calculation may be needed, or
a $T$-matrix has to be constructed using the phase shifts obtained from the DFT code and pair-interactions may need multiple scattering corrections. However,
one can learn from the corresponding $T=0$ problem where much of the
theory is available from e.g.,  Korringa, Kohn and
Rostoker~\cite{KKR}.  

In laser heating or shock compression, the theory assumes
that the external field has set up the $T_i$ and $T_e$ quasi-equilibrium. In
particular, the laser-field is assumed to be switched off. Hence the
method is independent of the laser intensity, polarization etc.
However, if a strong laser field is present,  then the DFT
calculation for the charge densities and transport properties have
to be carried out while including the self-consistent modification
of the occupation numbers of electronic states by the laser field,
as well as the associated dynamic screening of the electromagnetic
field, by extending the theory given in Refs.~\cite{zw,gri,lvm}
for strong electro-magnetic fields.
 Such calculations for WDM have not been attempted so far by anyone.

{\it Acknowledgment-} The author thanks Andrew Ng, and Michael
 Murrillo for their comments on an early version of the manuscript,
 while the help of Phil Stern and Damian Swift are acknowledged in 
 Ref.~\cite{Stern}.

\end{document}